# Inference and Prediction Using Functional Principal Components Analysis: Application to Diabetic Kidney Disease Progression in the Chronic Renal Insufficiency Cohort (CRIC) Study


Brian Kwan[1,2,a] , Wei Yang[3], Daniel Montemayor[4,5], Jing Zhang[2], Tobias Fuhrer[6,b], Amanda H. Anderson[7], Cheryl A.M. Anderson[8], Jing Chen[9], Ana C. Ricardo[10], Sylvia E. Rosas[11], Loki Natarajan[1,2], and the CRIC Study Investigators*

[1]Division of Biostatistics and Bioinformatics, Herbert Wertheim School of Public Health, University of California, San Diego, La Jolla, CA, USA;

[2]Moores Cancer Center, University of California, San Diego, La Jolla, CA, USA;

[3]Department of Biostatistics, Epidemiology and Informatics, Center for Clinical Epidemiology and Biostatistics, Perelman School of Medicine, University of Pennsylvania, Philadelphia, PA, USA;

[4]Division of Nephrology, Department of Medicine, University of Texas Health San Antonio, San Antonio, TX, USA;

[5]Center for Renal Precision Medicine, University of Texas Health San Antonio, San Antonio, TX, USA;

[6]Institute of Molecular Systems Biology, ETH Zurich, Zurich, Switzerland;

[7]Department of Epidemiology, Tulane School of Public Health and Tropical Medicine, New Orleans, LA;

[8]Herbert Wertheim School of Public Health, University of California, San Diego, La Jolla, CA, USA;

[9]Department of Medicine, Tulane School of Medicine, New Orleans, LA;

[10]Department of Medicine, University of Illinois, Chicago, IL, USA;

[11]Joslin Diabetes Center and Harvard Medical School, Boston, MA, USA;

*A list of the CRIC Study Investigators appears in the Acknowledgements.

**Correspondence:** Loki Natarajan, Division of Biostatistics, Herbert Wertheim School of Public Health, University of California San Diego, 3855 Health Sciences Dr #0901, La Jolla, CA 92093, USA. Email: lnatarajan@ucsd.edu.



Moved new affiliations:
[a] Department of Biostatistics, Fielding School of Public Health, University of California, Los Angeles, Los Angeles, CA, USA
[b] Biostarks, Geneva, Switzerland





**Abstract**

Repeated longitudinal measurements are commonly used to model long-term disease progression, and timing and number of assessments per patient may vary, leading to irregularly spaced and sparse data. Longitudinal trajectories may exhibit curvilinear patterns, in which mixed linear regression methods may fail to capture true trends in the data. We applied functional principal components analysis to model kidney disease progression via estimated glomerular filtration rate (eGFR) trajectories. In a cohort of 2641 participants with diabetes and up to 15 years of annual follow-up from the Chronic Renal Insufficiency Cohort (CRIC) study, we detected novel dominant modes of variation and patterns of diabetic kidney disease (DKD) progression among subgroups defined by the presence of albuminuria. We conducted inferential permutation tests to assess differences in longitudinal eGFR patterns between groups. To determine whether fitting a full cohort model or separate group-specific models is more optimal for modeling long-term trajectories, we evaluated model fit, using our goodness-of-fit procedure, and future prediction accuracy. Our findings indicated advantages for both modeling approaches in accomplishing different objectives. Beyond DKD, the methods described are applicable to other settings with longitudinally assessed biomarkers as indicators of disease progression. Supplementary materials for this article are available online.

**Keywords:** functional data analysis, longitudinal study, model fit, permutation tests, renal disease, sparse data






# 1       Introduction

Diabetes mellitus is the leading cause of chronic kidney disease (CKD) (Bailey et al. 2014; Centers for Disease Control and Prevention 2017; Koro et al. 2009; Koye et al. 2018; United States Renal Data System 2018). Patients with CKD are at increased risk of kidney failure, potentially requiring treatment by kidney transplant or dialysis. Yet, there is substantial heterogeneity in the development of kidney disease for patients with diabetes (Gheith et al. 2016).

Diabetic kidney disease (DKD) progression is typically characterized as the estimated glomerular filtration rate (eGFR) trajectory over time (de Boer et al. 2009; Robinson-Cohen et al. 2014). Linear mixed effect models are often used to estimate change in eGFR; however, characterizing the trajectory is complicated by observing sparse or irregularly spaced time series data which may exhibit nonlinear trends as depicted in <span style="color:red">Supplementary Figure 1</span> within the <span style="color:red">Supplementary Materials</span>. Thus, implementing flexible statistical methods that characterize eGFR trajectories in key subgroups could offer insights into DKD progression and treatment.

We propose the functional principal components analysis (FPCA) approach to model long-term trajectories while accounting for complexity in curve estimation, i.e., nonlinearity, sparsity, and irregularity. The subject of functional data analysis and the development of FPCA in particular is well studied (Ramsay and Silverman 2005; Wang et al. 2016) with applications of FPCA to sparse functional data dating back several decades (James et al. 2000; Paul and Peng 2009; Peng and Paul 2009; Rice and Wu 2001; Shi et al. 1996; Staniswalis and Lee 1998; Yao et al. 2005; Yao and Lee 2006). FPCA has also been applied to study various disease progression studies, e.g., lung, HIV (Szczesniak et al. 2017; Xie et al. 2017; Yan et al. 2017).



A salient feature of FPCA is that this approach can project infinite-dimensional curves into finite-dimensional vector scores and detect major modes of curve variation which would elucidate the leading patterns in disease (i.e., DKD) progression. Important for our context, expanding the FPCA framework to characterizing and predicting DKD progression by clinically meaningful subgroups will further elucidate the heterogeneous trends in the development of kidney disease across patients with diabetes. In this work, we will consider subgroups based on albuminuria, an established clinical biomarker for kidney disease based on the excess amount of albumin in the urine often used in conjunction with eGFR to classify patient CKD stage (Kidney Disease: Improving Global Outcomes (KDIGO) CKD Work Group 2013), although our approach could readily be applied to any other subgroups of interest. A key question in model development is whether a single *full cohort* FPCA model, trained using data from all patients with diabetes irrespective of albuminuria groups, is sufficient for characterizing and accurately predicting the long-term eGFR trajectory within specific albuminuria groups. If not, we consider multiple albuminuria *group-specific* models, each fitted using data from only patients of a specific albuminuria group, to prospectively predict eGFR trajectories for new subjects of the same group. To decide whether this group-level approach is needed, we examine differences in longitudinal eGFR patterns, i.e., mean and correlation functions, between albuminuria subgroups as well as model fit, via our proposed goodness-of-fit procedure, and future prediction accuracy. As a whole, our work applies and extends FPCA methodology to improving inference and prediction of trajectories of kidney function decline in heterogeneous subpopulations of patients with diabetes, with the methods generalizable to research applications beyond DKD.

The structure of this paper is as follows. Section 2 presents our methods. This describes the FPCA approach, inferential permutation tests to test for differences in longitudinal eGFR



patterns between groups, comparison of model fits and accuracy, our CRIC study cohort, assignment of albuminuria groups, and computational tools. Section 3 describes our statistical analysis results in detail. Lastly, Section 4 discusses the findings, limitations, and future directions of our work.

## 2 Methods

### 2.1 Functional Principal Components Analysis (FPCA)

Let $Y_{ij} = X_i(t_{ij}) + \epsilon_{ij}$ be the observed outcome at time $t_{ij}$, where $X_i(\cdot)$ is the measurement-error free outcome for subject $i$ at time $t_{ij}$ and $\epsilon_{ij}$ are measurement errors assumed to be identically and independently distributed normal with mean zero and variance $\sigma^2$ such that $i = 1, 2, \ldots, N$ and $j = 1, \ldots, m_i$.

We model the individual trajectories as a smooth random function $X(t)$ with unknown mean function $\mu(t)$ and covariance function $\Sigma(s, t)$, where $s, t \in \mathcal{T}$ and $\mathcal{T}$ is a bounded and closed time interval. Let $X_i(t)$ be the outcome trajectory for the $i$th individual and $t$ be years of follow-up. Under the Karhunen-Loeve expansion (Karhunen 1946; Loève 1946), the $i$th individual's trajectory can be expressed as $X_i(t) = \mu(t) + \sum_{k=1}^{\infty} \phi_k(t) * \xi_{ik}$ where $\phi_k(t)$ is the $k$th functional principal component (FPC) and $\xi_{ik}$ is the associated $k$th FPC score for the $i$th individual. The individual scores $\xi_{ik}$ are uncorrelated random variables with mean zero and variance $\lambda_k$, where $\lambda_1 \geq \lambda_2 \geq \cdots \geq 0$ and $\sum_k \lambda_k < \infty$. The covariance function $\Sigma(s, t)$ can be defined $\Sigma(s, t) = Cov(X_i(s) - \mu(s), X_i(t) - \mu(t)) = \sum_{k=1}^{\infty} \lambda_k * \phi_k(s) * \phi_k(t)$.

We briefly recapitulate the workflow and software implementation of the PACE algorithm by Yao et al. (2005) to estimate these model components in the <span style="color:red">Supplementary Materials</span>. Since the outcome trajectory is often well approximated by the top $K$ FPCs and their associated scores, we select $K$ as the number of FPCs that explained at least 95% of the total variance in the outcome



of interest. Compiling the above altogether for the PACE algorithm, we obtain the estimated

outcome trajectory for the $i$th individual as $\hat{X}_i(t) = \hat{\mu}(t) + \sum_{k=1}^{K} \hat{\phi}_k(t) * \hat{\xi}_{ik}$.

## 2.2    Testing for differences in mean and correlations functions

We used a permutation test based on a basis function representation to test for differences

in mean functions between $G$ groups by Górecki and Smaga (2015). The global null hypothesis

and its alterative are $H_0: \mu_1(t) = \mu_2(t) = \cdots = \mu_G(t)$ vs. $H_1: \exists\, u \neq v$ s.t. $\mu_u(t) \neq \mu_v(t)$,

respectively. To test for differences in correlation functions between $G$ groups, we center the

individual trajectories by the group mean trajectories and scale using the square root of the

diagonal of the smooth covariance estimates (standard deviations) from our full cohort model

and then apply the multiple-group permutation test developed by Cabassi et al. (2017) to test for

differences between the $G$ groups. The global null hypothesis and its alternative are $H_0: \Sigma_1 =$

$\Sigma_2 = \cdots = \Sigma_G$ vs. $H_1: \exists\, u \neq v$ s.t. $\Sigma_u \neq \Sigma_v$, respectively. Further details and the software

implementation for both permutation tests are provided in the <span style="color:red">Supplementary Materials</span>.

## 2.3    Evaluating model fit and accuracy

It is of interest to determine if the full cohort model, trained on all individuals irrespective

of subgroup (e.g., albuminuria category), is sufficient for predicting the long-term outcome

trajectory for individuals. If not, we consider separate group-specific models, fitted using only

individuals belonging to the same group, to prospectively predict group-specific outcome

trajectories for new individuals. To formally test whether the group-level approach is needed, we

develop a procedure for comparing the goodness-of-fit between our full cohort model and

multiple group-specific models via prediction error. In addition, we gauge the prediction

accuracy of future outcome values for individuals of different groups between our models.

### 2.3.1    Goodness-of-fit procedure



Our goodness-of-fit procedure, inspired by 5-fold cross-validation, compares the prediction error of our full cohort model and multiple group-specific models. Specifically for our full cohort model:

1) Divide the full cohort of individuals into 5 folds (4 for training and 1 for test data sets) of approximately equal size, ensuring that each fold contains approximately the same proportions of individuals in each group as that of the full cohort.

2) The model will predict the trajectories for individuals in the test set and we calculate the average curve squared error for the $i^{\text{th}}$ test individual as $ACSE_i = \sum_{j=1}^{n_i} \frac{[X_{ij} - \hat{X}_i(t_j)]^2}{n_i}$ where $n_i$ is the number of outcome observations for $i^{\text{th}}$ individual, $X_{ij}$ is the observed outcome for the $j^{\text{th}}$ observation of the $i^{\text{th}}$ individual, $\hat{X}_i(t_j)$ is the estimated outcome value at $t_j$ where $t_j$ is the time grid point closest in proximity to the individual's $j^{\text{th}}$ observation time. The mean average curve squared error $MACSE_d$ is then computed as the arithmetic mean of $ACSE_i$ values of the test set.

3) Repeat Step 2 for each of the other 4 folds as the test set. Our goodness-of-fit test statistic for the model (root $MACSE$) is Root $MACSE = \sqrt{\frac{1}{5} \sum_{d=1}^{5} MACSE_d}$.

The goodness-of-fit test statistic for the group-specific models will only involve training on the portion of individuals belonging to their respective groups across all 4 folds and the $MACSE_d$ is calculated for only the individuals belonging to the model's group in the test set. To estimate variability in the root $MACSE$, we repeat the above procedure to calculate 100 iterations of the goodness-of-fit statistic for the full cohort and group-specific models. Lower goodness-of-fit statistic values are indicative of better model fit.

2.3.2   Future Prediction Accuracy



We gauge the future outcome prediction accuracy of our models based on the mean squared error of individuals' latest observed outcome values and their corresponding model-based prediction. For the full cohort model, we train the model using data of all individuals, minus their latest outcome observation, and measure prediction accuracy for predicting this latest observation for the full cohort and various groups using mean squared error. The same approach is applied for the group-specific models, except training the models and calculating prediction accuracy only involves the portion of individuals belonging to their respective groups.

<u>2.4    Study cohort</u>

Our study cohort consisted of 2641 adults with diabetes and CKD enrolled in the Chronic Renal Insufficiency Cohort (CRIC) Study. Details on the rationale and design of the CRIC Study have been previously published (Denker et al. 2015; Feldman 2003; Kwan et al. 2020; Lash et al. 2009). To summarize, the CRIC Study recruited a racially and ethnically diverse patient population aged 21 to 74 years with varying CKD stages 2 (eGFR 60-70 ml/min/1.73m$^2$), 3a (eGFR 45-60 ml/min/1.73m$^2$), 3b (eGFR 30-45 ml/min/1.73m$^2$) and 4 (eGFR 20-30 ml/min/1.73m$^2$). Baseline characteristics of our CRIC study cohort is displayed in <span style="color:red">Supplementary Table 1</span> within the <span style="color:red">Supplementary Materials</span>.

Our main outcome is repeated measures of eGFR over time, calculated using the Chronic Kidney Disease Epidemiology Collaboration (CKD-EPI) creatinine equation (Levey et al. 2009). These measures were observed to be sparse and irregularly spaced, with a max follow-up median of 3.9 (range: 0-15.3, IQR: 2-7) years among participants. CRIC participants are assigned to albuminuria groups based on their urine albumin-creatinine ratio (ACR) values, normal to mildly increased (ACR < 30 mg/g), moderately increased (30-300 mg/g), and severely increased (> 300 mg/g), at baseline, the point of study entry.





All statistical analysis was conducted using the R (version 4.0.3) programming environment (R Core Team 2020).

## 3     Results

### 3.1     Full cohort FPCA

The full cohort (FPCA) model, fitted to our entire cohort (N=2641) of CRIC patients, estimated three leading FPCs were optimal, together capturing 98.5% of the variation in eGFR trajectories. These estimated FPCs and their associated scores are displayed in <span style="color:red">Figure 1</span>. The first FPC determined that 81.3% of the variation is explained by a magnitude shift from the full cohort model's mean eGFR trajectory. Thus, a large majority of the variation in eGFR trajectories can be traced to the patients' measured eGFR at study entry, which lends to the idea that the first FPC behaves like a model intercept. The first FPC scores significantly differed by albuminuria group both globally and pairwise (all $p < 0.001$). Patients in the normal-mild group tended to have higher scores and a greater proportion of patients in this group have positive valued scores than the other two groups. Hence, resting solely on the first FPC, patients with normal-mild albuminuria are more inclined to have consistently increased eGFR compared to patients with moderate or severe albuminuria. A notable proportion of patients with severe albuminuria have negative valued first FPC scores along with mostly lower scores than those with normal-mild and moderate albuminuria. Accordingly, the first FPC exerted that patients with severe albuminuria are more inclined to have consistently decreased eGFR.

The second FPC accounted for 13.5% of the variation and captured varying rates of change in eGFR in the first and second halves of our follow-up time frame. Similar to how the first FPC is a surrogate to the model intercept, the second FPC is akin to that of the model slope.



The second FPC scores significantly differed by albuminuria group both globally and pairwise (all $p < 0.001$, except normal-mild vs moderate $p = 0.013$). Here, the distribution of the second FPC scores for the normal-mild and moderate groups are alike with one another, while the patients with severe albuminuria tended to have slightly lower scores. Therefore, the rate of change in eGFR for patients with severe albuminuria are more differentiated from those with normal-mild or moderate albuminuria. Finally, the third FPC explained 3.7% of the variation and also captured varying rates of change in eGFR although with different inflection points. The third FPC scores significantly differed by albuminuria group both globally and pairwise (all $p < 0.001$). Much like the second FPC scores, the distribution of the scores here for the normal-mild and moderate groups are more alike with one another. In contrast to before, now the patients with severe albuminuria tended to have higher third FPC scores than those with normal-mild or moderate albuminuria, although the difference is less notable compared to the second FPC scores comparison, which can be indicative of survival bias for the severe group with slower kidney function decline toward the end of the follow-up time frame. Taking both the second and third FPCs and their scores into account, the eGFR trajectories of patients with severe albuminuria are more prone to having various rate of changes throughout the follow-up period than those with normal-mild or moderate albuminuria.

*Comparing mean trajectories:* The estimated eGFR trajectories for all the CRIC patients from the full cohort model are illustrated in <span style="color:red">Figure 2</span>. The mean eGFR trajectories for the albuminuria groups were calculated by taking the sample means of the FPCA-estimated eGFR at each time grid point (years) of the group-specific patients. By Górecki and Smaga's permutation test (Górecki and Smaga 2015), the mean eGFR trajectories for the albuminuria groups differed both globally (FP = 318.86, $p < 0.001$) and pairwise (normal-mild vs moderate FP = 74.13,



normal-mild vs severe FP = 641.34, moderate vs severe FP = 239.27, all p < 0.001). The normal-mild group's mean eGFR trajectory was consistently greater than those of the other groups by an increased magnitude shift, largely attributed to the first FPC and the associated scores for the patients in the normal-mild group being notably higher than patients in the moderate or severe groups. The moderate group's mean eGFR trajectory, although a downward shift from that of the normal-mild group, had a similar rate of decline over time. As mentioned before, the second and third FPCs captured the varying rates of change in eGFR for our full cohort and both their associated scores are extremely similar in distribution for the normal-mild and moderate groups, which led to similar observed rates of change in eGFR for both groups. The severe group's mean eGFR trajectory differed from the other group's mean trajectories in both magnitude shift and rates of change. In particular, the trajectory has a steeper initial decline than the normal-mild and moderate groups and a noticeable rebound of the trajectory for later follow-up period is owed to both the second and third FPCs. Here, the second FPC was negative along much of the same late follow-up period as the eGFR rebound, and patients with severe albuminuria having more negative valued scores, contributed to a positive rate of change in eGFR. The third FPC was positive for the similar follow-up period as the eGFR rebound and the patients with severe albuminuria having more positive valued scores further pitched in to the increase in eGFR towards the end of follow-up. However, we interpret this with caution due to sparsity of samples in the severe group in later years: the severe group had < 50 eGFR samples after 11.84 years, which is earlier than that of normal-mild and moderate groups (13.92 and 12.99 years, respectively).

    *Comparing correlation functions:* The estimated correlation functions of the eGFR trajectories for each albuminuria group are presented in Figure 3. The correlation at any two time

points is not lower than 0.55 for any group. By Cabassi et al.'s permutation test (Cabassi et al. 2017), the correlation functions of eGFR for the albuminuria groups differed both globally (p < 0.001) and pairwise (normal-mild vs moderate p = 0.028, normal-mild vs severe p < 0.001, moderate vs severe p < 0.001), although accounting for multiple comparisons via a Bonferroni correction, the normal-mild and moderate groups would not differ. We observed that the severe group contained a wider area of correlations < 0.7 than the other groups. This particular area is largely concentrated on the correlations between eGFR at the earlier time period < 5 years and that of the later time period > 5 years. Within this area, the severe group correlations between baseline (year 0), and > 5 years was mostly lower than those of the normal-mild and moderate groups, indicated by values < 0.65 in the dark blue region, which can be attributed to the faster decline of eGFR post 5 years of follow-up in the group characterize as severe.

## 3.2    Group-specific FPCA

The albuminuria-specific (group) models, each fitted using only data from CRIC patients of a particular albuminuria group, estimated varying leading FPCs for capturing at least 95% of the variation in the eGFR trajectories of each group and are displayed in Figure 4. In addition, the same estimated FPCs from the full cohort model, displayed in panel (a) of Figure 1, are overlaid on these plots for comparison with those of the group models. The first FPC for the normal-mild model was similar to that of the full cohort model; however, the moderate and severe models did not have as consistent a magnitude shift over time from their mean eGFR trajectories. In fact, the first FPCs for the moderate and severe models did not account for as much variation in the patient trajectories (< 80%) and have asymptotic value 0 with respect to follow-up time, which could simply reflect fewer samples at later follow-up. Inversely, the second FPC for these models explained more variation (> 17%) compared to the full cohort and



normal-mild models ($< 14\%$). The second FPCs for the four models captured varying rates of eGFR change, although those of the moderate and severe models closely resemble reflections of the full cohort and normal-mild models about the value 0, with the severe model having a more steady rate of change towards the last few years of follow-up, again reflecting sparse data in this group at later follow-up years. Finally, we note (Figure 4) that the normal-mild model only estimated two leading FPCs while the other models contained three, with the severe model's third FPC undergoing a sharp decline near the last years of follow-up. As a preliminary visual assessment of model fit, we compared the mean and randomly selected individual fitted eGFR trajectories for the full cohort and group-specific models in Supplementary Figure 2 within the Supplementary Materials. However, these results may be over optimistic, since these patients were also used to train the models. Thus, we investigated cross-validated prediction error using the goodness-of-fit statistic as outlined in Section 2.3.1.

*Evaluating model fit:* The results of 100 iterations of our proposed goodness-of-fit procedure for comparing between the single full cohort model and three albuminuria-specific models are presented in Figure 5. Our goodness-of-fit procedure is based on prediction error between observed and predicted eGFR, with lower root mean average curve squared error (MACSE) values indicative of better model fit. The full cohort model displayed largely better prediction performance for patients with normal-mild albuminuria such that there is a lower distribution of cross-validated root MACSE values than using the normal-mild model. For patients with moderate albuminuria, the full cohort model had comparable prediction to the moderate model. In contrast, the severe model had better prediction performance for patients with severe albuminuria with a lower distribution of cross-validated root MACSE values than using the full cohort model. Furthermore, this is the largest gap in prediction performance



between the full cohort and an albuminuria-specific model. The range of root MACSE estimates across all three groups of patients in using the single full cohort model [4.61-5.61] was less than using three albuminuria-specific models [3.97-5.91]. Finally, in gauging future prediction accuracy of these models by comparing between latest observed and predicted eGFR, the full cohort model had a root MSE values of 10.81 for full cohort, 9.82 for the normal to mild group, 10.14 for the moderate group, and 12.47 for the severe group. Meanwhile, the group-specific normal-mild, moderate, and severe models had root MSE values of 9.74, 10.09, and 10.8, respectively. Therefore, the group-specific models performed marginally better in future prediction. Most of these models had an approximate error of 10 ml/min/1.73m$^2$ on average between latest observed and predicted eGFR with the exception of the full cohort model's prediction for the severe group having an approximate error of 12 ml/min/1.73m$^2$.

## 4     Discussion

We applied FPCA methodology to model long-term eGFR trajectories for patients with diabetes and CKD that accounted for nonlinear, sparse, and irregularly spaced eGFR time series. The first two leading FPCs, accounted for 95% of the variation in eGFR patterns, and behaved similarly to a model's intercept and slope respectively. As an innovative application, we implemented functional data inferential methods for comparing mean and correlation functions to assess whether the longitudinal eGFR patterns significantly varied by level of albuminuria. Dong et al. (2018) note the need to develop methods for predicting future eGFR for patients of different clinical subgroups. One of their proposed solutions to this is fitting separate FPCA for each subgroup much like our group-level approach. However, we further extend this idea by incorporating a goodness-of-fit procedure that uses cross-validated leave-a-curve-out prediction



errors to decide between fitting separate albuminuria group-specific FPCA models and using a single full cohort FPCA model.

We found that the most dominant mode of eGFR variation was a magnitude shift from the mean eGFR for all groups whether fitting a full cohort or group model, having explained > 70% of the variability. Lesser dominant modes involved varying rates and time-intervals of eGFR change; these modes were pertinent for modeling trajectories for patients in the moderate and severe groups than those in the normal-mild group, with the second FPCs explaining > 17% of the eGFR variation for the moderate and severe groups versus ~9% for the normal-mild group. Of interest, a third FPC was not required for the normal-mild group, in contrast to the moderate and severe groups, in which the third FPC explained 4% of the variation in eGFR trends. Mean and correlation eGFR functions significantly differed between albuminuria groups, especially the severe versus the other two groups. Upon further inspection, we found that the correlations between eGFR at the earlier time period of < 5 years and that of the later time period > 5 years are noticeably lower for the severe group, suggesting more diffuse, i.e., less tightly linked long-term eGFR trajectories in patients with more severe kidney disease. Results from our goodness-of-fit procedure of prediction performance indicated that the choice of a single full cohort model is a viable option to predicting long-term eGFR trajectories for different albuminuria groups. In particular, although the moderate and severe specific models had lower prediction error compared to the full cohort model, these group-specific models also had more variability in prediction error suggesting the classic tradeoff between low bias (group-specific models) versus low variance (full cohort model).

Key methodological strengths of our work include the application of FPCA, which models non-linear trajectories, and investigates leading modes of variation, while overcoming



sparsity and irregularly spaced trends. This approach, thus permits a nuanced assessment of long-term disease progression as evidenced in our findings. In addition, inference for mean and correlation functions between groups were based on permutation tests and are therefore robust to misspecification of distribution. Our goodness-of-fit procedure gauges the variability of curve prediction error via repeated cross-validated measures which reduces overfitting and avoids optimistic prediction performance for the full cohort and group-specific models. An important epidemiological strength is the use of the CRIC study, one of the largest cohorts of individuals with diabetes in the U.S., with extensive clinical profiles and extended longitudinal data of kidney function.

From a statistical viewpoint, a limitation is that data were sparse at later follow-up times, especially for the severe albuminuria group, hence any conclusions for later times need to be interpreted with caution. Although our proposed goodness-of-fit procedure for FPCA model comparison in estimating long-term outcome trajectories is relatively novel, more work in this area is needed. We aim to further its methodological developments. For instance, to investigate the precision of our model-driven estimate for prediction error, we would like to explore bootstrap resampling techniques for deriving an empirical distribution of the standard error of our estimates, rather than using cross-validated estimates which could suffer from small test sample-size. Also, our FPCA models do not take into account the possibility that the eGFR time series data could be subjected to informative missingness, which could result in biased estimation of our functional principal components. While Shi et al. (2021) recently developed an extension of FPCA well suited for handling informative missingness, we note that our group-specific model approach provided separate estimation of mean and covariance functions allowing for distinct inference and prediction of longitudinal DKD for each albuminuria group



based on their group-specific missing data patterns. In addition, our statistical analysis was conducted on the diverse longitudinal CRIC study cohort of 2641 patients with diabetes in which the number of patients for each albuminuria group (normal-mild: 965, moderate: 768, severe: 908) are each individually greater than the total number of patients from their study (N=252) as well as having a greater follow-up period (15 > 10 years).

Within the kidney disease setting, Dong et al. (2018) applied FPCA to uncover major sources of variations of glomerular filtration rate trajectories of kidney transplant recipients. Our work is different and expands on this previous study's statistical approaches by (1) centering the application on modeling and investigating salient patterns of eGFR trajectories of patients with diabetes and CKD and (2) incorporating statistical inference and prediction methods to investigate for differences in DKD progression between key clinical subgroups from FPCA model development. Our first two leading FPCs essentially mimic intercept (i.e., starting eGFR value) and slope (i.e. linear rate of change of eGFR), suggesting that standard statistical approaches such as linear mixed models would be sufficient for capturing the majority of variation in kidney disease patterns. Nevertheless, the more subtle variations captured by the third FPC, albeit only explaining 4% of the variation, may still be important for individual patients, especially those with more advanced albuminuria stage. Similarly, while the full cohort model fit the data well and had low prediction error overall, there were differences in performance based on albuminuria severity, with group-specific models being more advantageous in the severe albuminuria group. The decision of training a single model over multiple models to model long-term trajectories is a question of parsimony versus complexity, i.e., the full cohort model is simpler, computationally less demanding, and more generalizable since it is applied to the entire spectrum of albuminuria. Thus, a single FPCA model, or even a



more traditional intercept and slope model, can be recommended in population settings where the goal is to model population level trends, and test broad brush differences between groups. However, in a clinical setting group-specific models may offer better predictive performance and offer the opportunity for personalized recommendations for individuals presenting with more advanced disease. In summary, in this work we leverage the vast and established FPCA theory to provide innovative analytic tools and practical data analysis strategies for modeling disease progression and informing risk stratification in clinical settings.

## Acknowledgements


The CRIC Study investigators consisted of Lawrence J. Appel, MD, MPH, Debbie L. Cohen, MD, Harold I. Feldman, MD, MSCE, Alan S. Go, MD, James P. Lash, MD, Robert G. Nelson, MD, PhD, MS, Mahboob Rahman, MD, Panduranga S. Rao, MD, Vallabh O. Shah, PhD, MS, Mark L. Unruh, MD, MS. We thank Lisa E. Wesby, MS for her assistance as CRIC project manager in managing and facilitating the manuscript among all the authors. This material is based upon work supported by the National Science Foundation Graduate Research Fellowship Program under Grant No. DGE-1650112. Any opinions, findings, and conclusions or recommendations expressed in this material are those of the author(s) and do not necessarily reflect the views of the National Science Foundation.


## Supplementary Materials

Our Supplemental Materials (later below) consists of the supplementary figures and table as well as details and software implementation of our statistical methods.

## References


Bailey, R. A., Wang, Y., Zhu, V., and Rupnow, M. F. (2014), "Chronic kidney disease in US adults with type 2 diabetes: An updated national estimate of prevalence based on Kidney





Disease: Improving Global Outcomes (KDIGO) staging," *BMC Research Notes*, 7, 1–7.

de Boer, I. H., Katz, R., Fried, L. F., Ix, J. H., Luchsinger, J., Sarnak, M. J., Shlipak, M. G., Siscovick, D. S., and Kestenbaum, B. (2009), "Obesity and Change in Estimated GFR Among Older Adults," *American Journal of Kidney Diseases*, Elsevier Inc., 54, 1043–1051.

Cabassi, A., Pigoli, D., Secchi, P., and Carter, P. A. (2017), "Permutation tests for the equality of covariance operators of functional data with applications to evolutionary biology," *Electronic Journal of Statistics*, 11, 3815–3840.

Centers for Disease Control and Prevention (2017), *National Chronic Kidney Disease Fact Sheet, 2017*, Atlanta, GA.

Denker, M., Boyle, S., Anderson, A. H., Appel, L. J., Chen, J., Fink, J. C., Flack, J., Go, A. S., Horwitz, E., Hsu, C., Kusek, J. W., Lash, J. P., Navaneethan, S., Ojo, A. O., Rahman, M., Steigerwalt, S. P., Townsend, R. R., and Feldman, H. I. (2015), "Chronic Renal Insufficiency Cohort Study (CRIC): Overview and Summary of Selected Findings," *Clinical Journal of the American Society of Nephrology*, 10, 2073–2083.

Dong, J. J., Wang, L., Gill, J., and Cao, J. (2018), "Functional principal component analysis of glomerular filtration rate curves after kidney transplant," *Statistical Methods in Medical Research*, 27, 3785–3796.

Feldman, H. I. (2003), "The Chronic Renal Insufficiency Cohort (CRIC) Study: Design and Methods," *Journal of the American Society of Nephrology*, 14, 148S – 153.

Gheith, O., Farouk, N., Nampoory, N., Halim, M. A., and Al-Otaibi, T. (2016), "Diabetic kidney disease: world wide difference of prevalence and risk factors.," *Journal of nephropharmacology*, 5, 49–56.

Górecki, T., and Smaga, Ł. (2015), "A comparison of tests for the one-way ANOVA problem for



functional data," *Computational Statistics*, Springer Berlin Heidelberg, 30, 987–1010.

James, G. M., Hastie, T. J., and Sugar, C. A. (2000), "Principal component models for sparse functional data," *Biometrika*, 87, 587–602.

Karhunen, K. (1946), "Zur Spektraltheorie stochastischer Prozesse," *Ann. Acad. Sci. Fennicae, AI*, 34.

Kidney Disease: Improving Global Outcomes (KDIGO) CKD Work Group (2013), *KDIGO 2012 Clinical Practice Guideline for the Evaluation and Management of Chronic Kidney Disease*.

Koro, C. E., Lee, B. H., and Bowlin, S. J. (2009), "Antidiabetic medication use and prevalence of chronic kidney disease among patients with type 2 diabetes mellitus in the United States," *Clinical Therapeutics*, Excerpta Medica Inc., 31, 2608–2617.

Koye, D. N., Magliano, D. J., Nelson, R. G., and Pavkov, M. E. (2018), "The Global Epidemiology of Diabetes and Kidney Disease," *Advances in Chronic Kidney Disease*, Elsevier Ltd, 25, 121–132.

Kwan, B., Fuhrer, T., Zhang, J., Darshi, M., Van Espen, B., Montemayor, D., de Boer, I. H., Dobre, M., Hsu, C. yuan, Kelly, T. N., Raj, D. S., Rao, P. S., Saraf, S. L., Scialla, J., Waikar, S. S., Sharma, K., Natarajan, L., Appel, L. J., Feldman, H. I., Go, A. S., He, J., Lash, J. P., Rahman, M., and Townsend, R. R. (2020), "Metabolomic Markers of Kidney Function Decline in Patients With Diabetes: Evidence From the Chronic Renal Insufficiency Cohort (CRIC) Study," *American Journal of Kidney Diseases*, 76, 511–520.

Lash, J. P., Go, A. S., Appel, L. J., He, J., Ojo, A., Rahman, M., Townsend, R. R., Xie, D., Cifelli, D., Cohan, J., Fink, J. C., Fischer, M. J., Gadegbeku, C., Hamm, L. L., Kusek, J. W., Landis, J. R., Narva, A., Robinson, N., Teal, V., and Feldman, H. I. (2009), "Chronic Renal



Insufficiency Cohort (CRIC) Study: Baseline Characteristics and Associations with Kidney Function," *Clinical Journal of the American Society of Nephrology*, 4, 1302–1311.

Levey, A. S., Stevens, L. A., Schmid, C. H., Zhang, Y., Castro, A. F., Feldman, H. I., Kusek, J. W., Eggers, P., Lente, F. Van, Greene, T., and Coresh, J. (2009), "A new equation to estimate glomerular filtration rate," *Annals of Internal Medicine*, 150, 604–612.

Loève, M. (1946), "Fonctions aléatoires à décomposition orthogonale exponentielle," *La Revue Scientifique*, 84, 159–162.

Paul, D., and Peng, J. (2009), "Consistency of restricted maximum likelihood estimators of principal components," *Annals of Statistics*, 37, 1229–1271. https://doi.org/10.1214/08-AOS608.

Peng, J., and Paul, D. (2009), "A geometric approach to maximum likelihood estimation of the functional principal components from sparse longitudinal data," *Journal of Computational and Graphical Statistics*, 18, 995–1015.

R Core Team (2020), "R: A Language and Environment for Statistical Computing," Vienna, Austria: R Foundation for Statistical Computing.

Ramsay, J. O., and Silverman, B. W. (2005), *Functional Data Analysis*, Springer Series in Statistics, New York, NY: Springer New York.

Rice, J. A., and Wu, C. O. (2001), "Unequally Sampled Noisy Curves," *Biometrics*, 57, 253–259.

Robinson-Cohen, C., Littman, A. J., Duncan, G. E., Weiss, N. S., Sachs, M. C., Ruzinski, J., Kundzins, J., Rock, D., De Boer, I. H., Ikizler, T. A., Himmelfarb, J., and Kestenbaum, B. R. (2014), "Physical activity and change in estimated GFR among persons with CKD," *Journal of the American Society of Nephrology*, 25, 399–406.

Shi, H., Dong, J., Wang, L., and Cao, J. (2021), "Functional principal component analysis for



longitudinal data with informative dropout," *Statistics in Medicine*, 40, 712–724.

Shi, M., Weiss, R. E., and Taylor, J. M. G. (1996), "An Analysis of Paediatric CD4 Counts for Acquired Immune Deficiency Syndrome Using Flexible Random Curves," *Applied Statistics*, 45, 151.

Staniswalis, J. G., and Lee, J. J. (1998), "Nonparametric Regression Analysis of Longitudinal Data," *Journal of the American Statistical Association*, 93, 1403–1418.

Szczesniak, R. D., Li, D., Su, W., Brokamp, C., Pestian, J., Seid, M., and Clancy, J. P. (2017), "Phenotypes of rapid cystic fibrosis lung disease progression during adolescence and young adulthood," *American Journal of Respiratory and Critical Care Medicine*, 196, 471–478.

United States Renal Data System (2018), *2018 USRDS Annual Data Report: Epidemiology of kidney disease in the United States*, Bethesda, MD.

Wang, J.-L., Chiou, J.-M., and Müller, H.-G. (2016), "Functional Data Analysis," *Annual Review of Statistics and Its Application*, 3, 257–295.

Xie, W., Agniel, D., Shevchenko, A., Malov, S. V., Svitin, A., Cherkasov, N., Baum, M. K., Campa, A., Gaseitsiwe, S., Bussmann, H., Makhema, J., Marlink, R., Novitsky, V., Lee, T. H., Cai, T., O'Brien, S. J., and Essex, M. (2017), "Genome-Wide Analyses Reveal Gene Influence on HIV Disease Progression and HIV-1C Acquisition in Southern Africa," *AIDS Research and Human Retroviruses*, 33, 596–609.

Yan, F., Lin, X., and Huang, X. (2017), "Dynamic prediction of disease progression for leukemia patients by functional principal component analysis of longitudinal expression levels of an oncogene," *Annals of Applied Statistics*, 11, 1649–1670.

Yao, F., and Lee, T. C. M. (2006), "Penalized spline models for functional principal component analysis," *Journal of the Royal Statistical Society. Series B: Statistical Methodology*, 68, 3–




25.


Yao, F., Müller, H. G., and Wang, J. L. (2005), "Functional data analysis for sparse longitudinal data," *Journal of the American Statistical Association*, 100, 577–590.




<u>Figure 1:</u> (a) Leading three FPCs for our full cohort model with proportion of eGFR variance explained (PVE %). (b) Box plots of the scores for the leading FPCs by albuminuria group. The three FPC scores significantly differed by albuminuria group both globally and pairwise (all p < 0.001, except normal-mild vs moderate for FPC 2 scores has p = 0.013).
Albuminuria group:  NoMi: Normal-Mild, Mod: Moderate, Sev: Severe

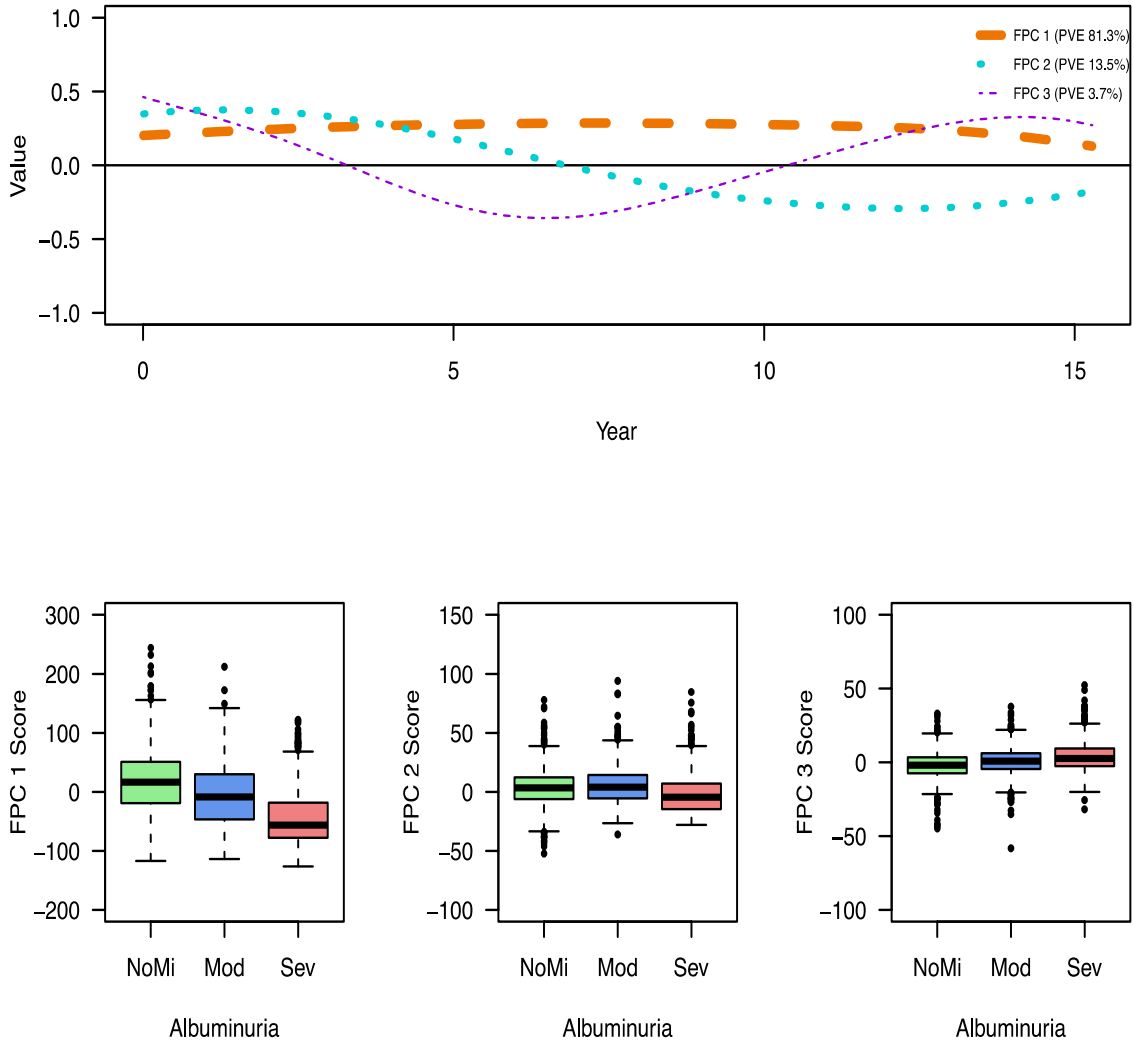



<u>Figure 2:</u> Estimated individual eGFR trajectories from our full cohort model. The thick curves are the albuminuria group mean trajectories with a black diamond corresponding to when that group has < 50 eGFR samples after that time point (Normal-Mild – 13.92 years, Moderate – 12.99 years, Severe – 11.84 years).

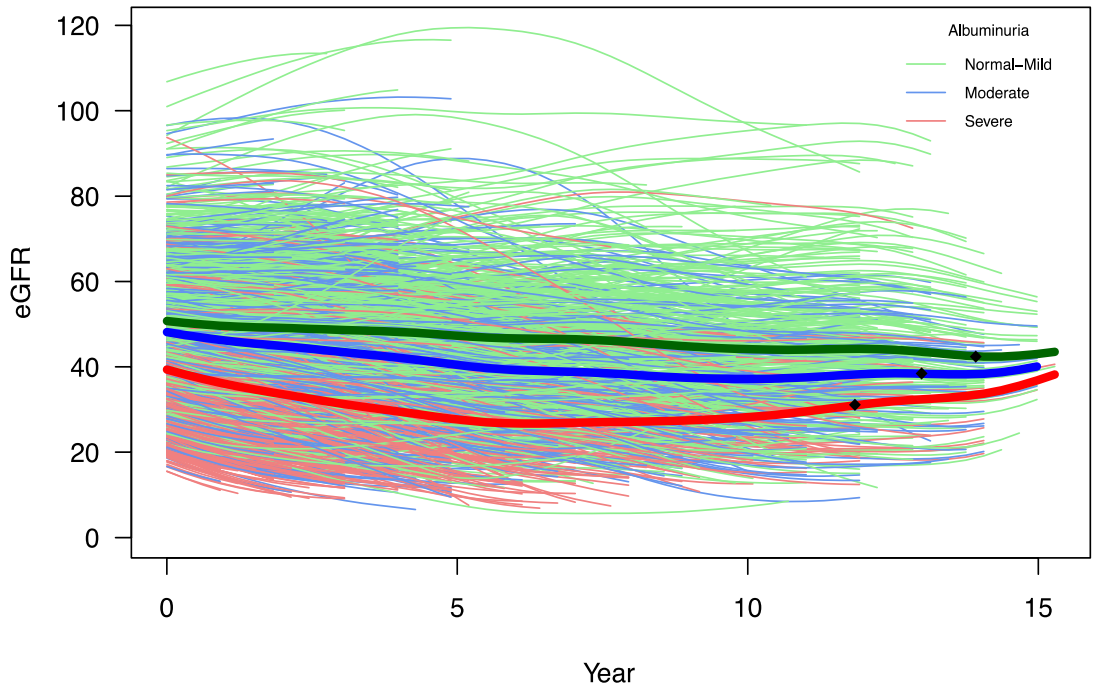



**Figure 3:** Estimated correlation functions for the normal-mild, moderate, and severe albuminuria groups.

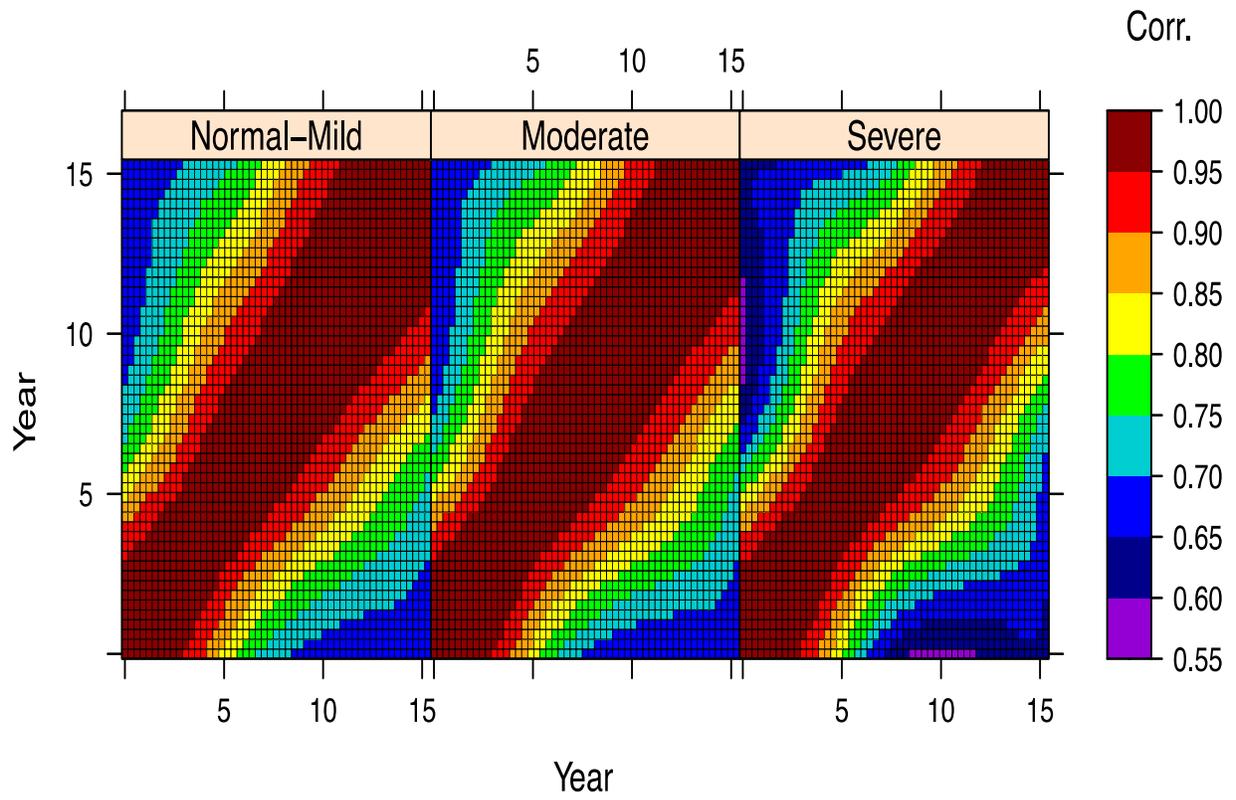



Figure 4: Leading three FPCs for our full cohort and group-specific models along with proportion of eGFR variance explained (PVE %). Notably, the normal-mild model only required the leading two FPCs to explain at least 95% of the total eGFR variation; therefore, only the leading two FPCs for this model are plotted.

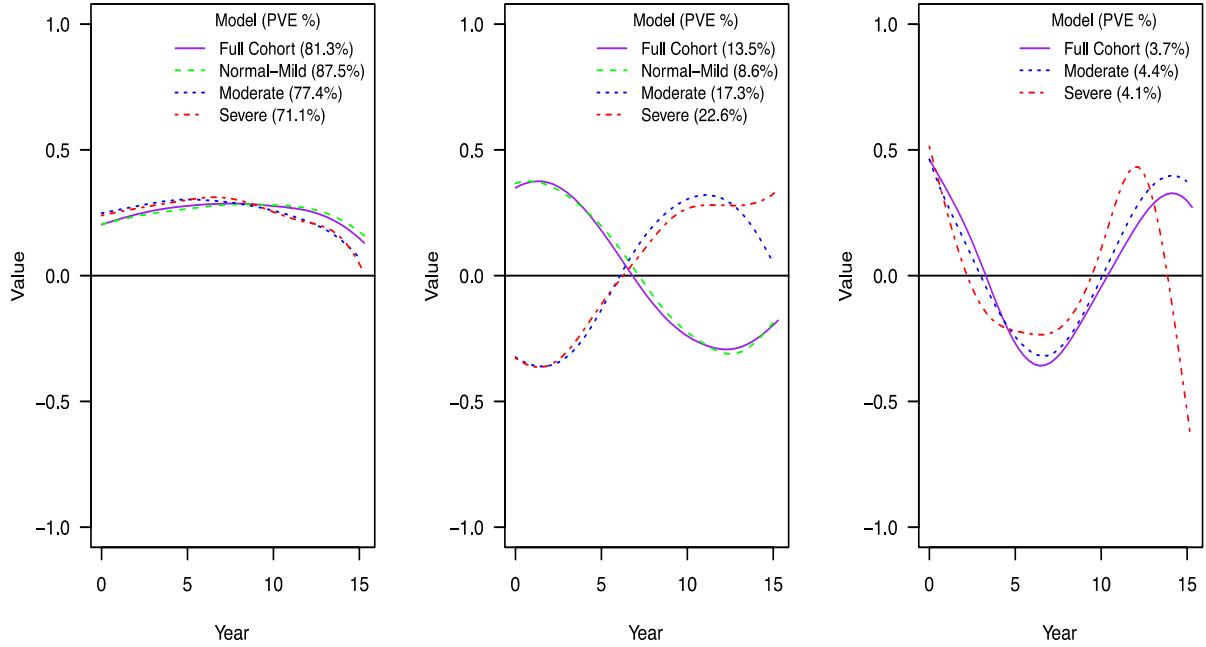



<u>Figure 5:</u> Box plots of model prediction performance: 100 repeats of 5-fold cross-validated root MACSE (mean average curve squared error) for FPCA-estimated eGFR trajectories.

- NoMi: Normal-Mild, Mod: Moderate, Sev: Severe
- FC, Full Cohort; FC-NoMi, FC-Mod, FC-Sev; Full Cohort prediction for the respective group-specific patients.

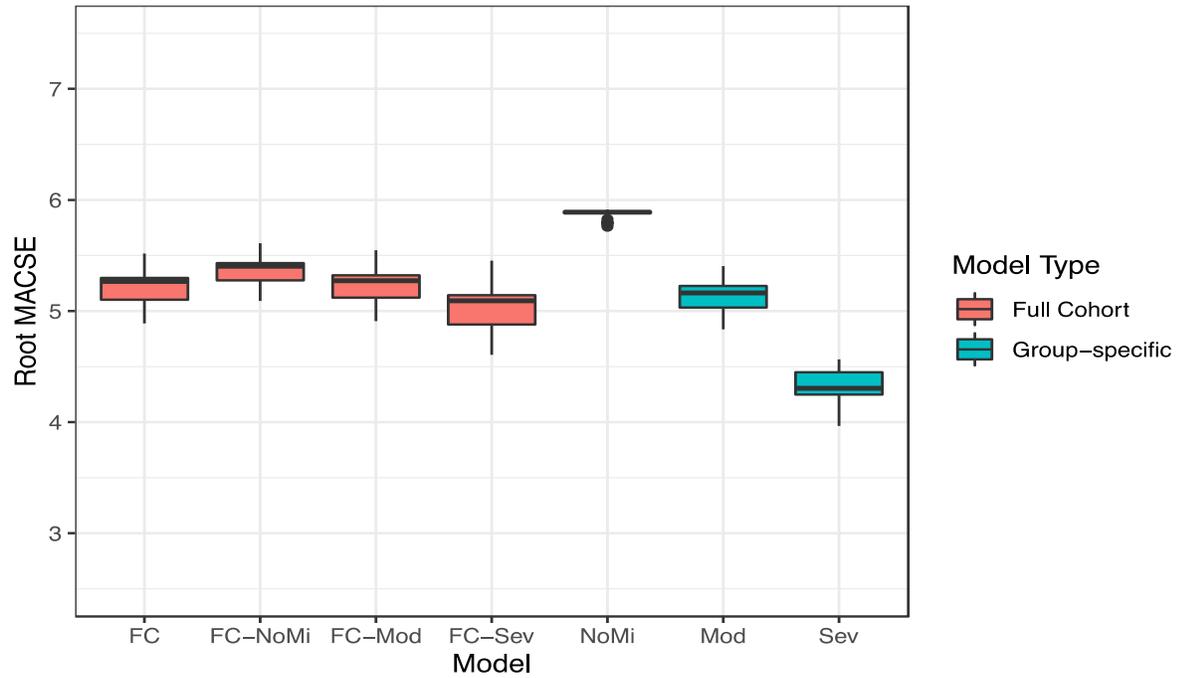



Supplementary Materials
for
**Inference and Prediction Using Functional Principal Components Analysis: Application to Diabetic Kidney Disease Progression in the Chronic Renal Insufficiency Cohort (CRIC) Study**

by Brian Kwan, Wei Yang, Daniel Montemayor, Jing Zhang, Tobias Fuhrer, Amanda H. Anderson, Cheryl A.M. Anderson, Jing Chen, Ana C. Ricardo, Sylvia E. Rosas, Loki Natarajan, and the CRIC Study Investigators

Supplementary Figure 1: Various patterns of observed eGFR trajectories for our study sample of patients with diabetes in the Chronic Renal Insufficiency Cohort (CRIC) study, including sparse or irregular spaced data.

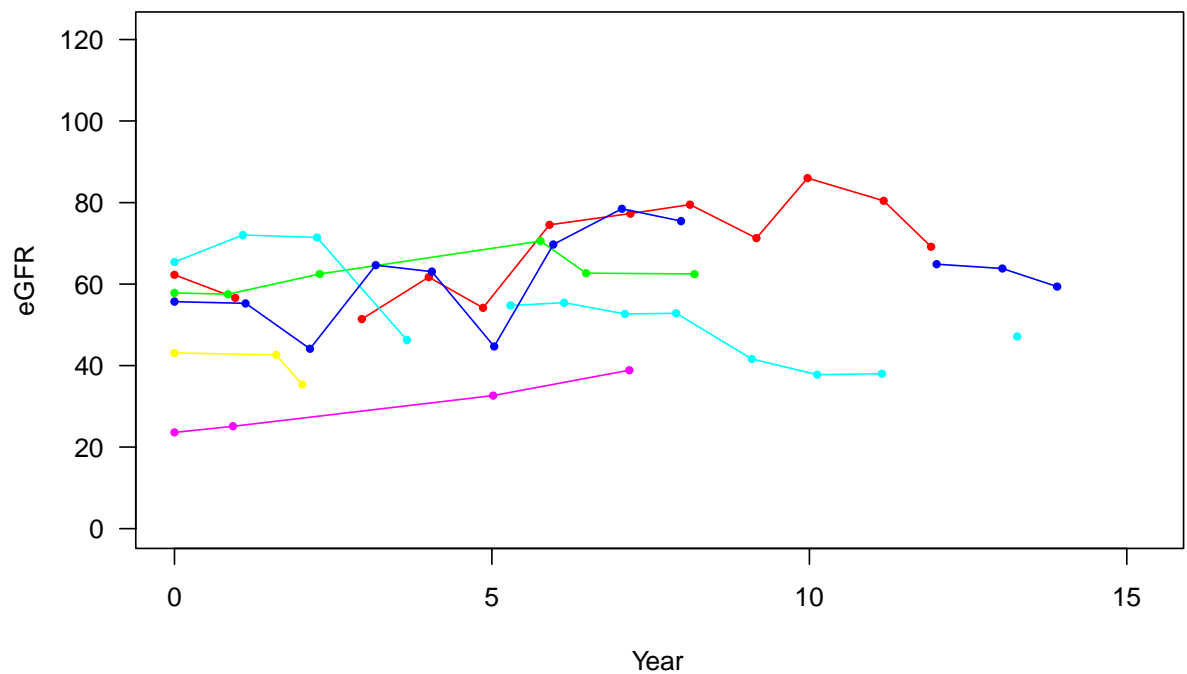



<u>Estimating the model components of FPCA using the PACE algorithm by Yao et al. (2005)</u>

The mean function $\hat{\mu}(t)$ is estimated using a local linear smoother that aggregates data from all individuals. The smooth covariance is estimated from the individual "raw" covariances, $\Sigma_i(t_{ij}, t_{il}) = (Y_{ij} - \hat{\mu}(t_{ij}))(Y_{il} - \hat{\mu}(t_{il}))$ the $i$th individual's raw covariance. The estimates of the FPCs (eigenfunctions) $\hat{\phi}_k$ and its eigenvalues $\hat{\lambda}_k$ are solutions to the eigenequations $\int_{\mathcal{T}} \hat{\Sigma}(s,t)\hat{\phi}_k(s)ds = \hat{\lambda}_k\hat{\phi}_k(t)$ satisfying the constraints $\int_{\mathcal{T}} \hat{\phi}_k(t)^2 dt = 1$ and $\int_{\mathcal{T}} \hat{\phi}_k(t)\hat{\phi}_m(t)dt = 0$ for $m < k$. Finally, the estimated $k$th FPC score for the $i$th individual is acquired through conditional expectation $\hat{\xi}_{ik} = \hat{E}(\xi_{ik}|Y_i) = \hat{\lambda}_k\hat{\phi}_{ik}^T\hat{\Sigma}_{Y_i}^{-1}(Y_i - \hat{\mu}_i)$ such that $Y_i = (Y_{i1}, \dots, Y_{im_i})^T$. The components $\hat{\phi}_{ik}$ and $\hat{\mu}_i$ are vector variants of $\hat{\phi}_k(t)$ and $\hat{\mu}(t)$, respectively, evaluated on a grid of time points $t_{ij}, j = 1, \dots, m_i$.

The R package fdapace was used for the estimation of the model components of FPCA.

<u>Assessing group differences in mean functions using the test by Górecki and Smaga (2015)</u>

Suppose that we have individual smooth random functions $X_{gi} \in L_2(\mathcal{T})$ indexed by $G$ groups, where $g = 1, \dots, G$ and $i = 1, \dots, n_g$ such that $\sum_{g=1}^{G} n_g = N$. The trajectories $X_{gi}(t)$ can be represented as a linear combination of orthonormal basis functions $\{\psi_l\}$ of $L_2(\mathcal{T})$, $X_{gi}(t) = \sum_{l=1}^{L} c_{gil} * \psi_l(t)$ where $t \in \mathcal{T}$ and $c_{gil}$ are random variables with finite variance. Defining $\boldsymbol{\psi}(t) = (\psi_0(t), \ \psi_1(t), \dots, \psi_L(t))^T$ and $\boldsymbol{c}_{gi} = (c_{gi0}, c_{gi1}, \dots, c_{giL})^T$, we can represent the individual, sample group mean, and sample grand mean trajectories as $X_{gi}(t) = \boldsymbol{c}'_{gi}\boldsymbol{\psi}(t)$, $\bar{X}_g(t) = \frac{1}{n_g}\sum_{i=1}^{n_g} \boldsymbol{c}'_{gi}\boldsymbol{\psi}(t)$, $\bar{X}(t) = \frac{1}{N}\sum_{g=1}^{G}\sum_{i=1}^{n_g} \boldsymbol{c}'_{gi}\boldsymbol{\psi}(t)$, respectively. The F-test statistic for



the one-way ANOVA problem can be adapted for the functional data setting as $FP =$

$$\frac{\frac{1}{G-1}\sum_{g=1}^{G}n_g||\bar{X}_g - \bar{X}||_2^2}{\frac{1}{N-G}\sum_{g=1}^{G}\sum_{i=1}^{n_g}||X_{gi} - \bar{X}_g||_2^2} \text{ where } ||f||_2^2 = \int_{\mathcal{T}} f^2(t)dt \text{ for } f \in L_2(\mathcal{T}).$$

Here, $FP$ serves as the test statistic for a permutation-based p-value in testing the global null hypothesis. We set 1000 permutation replicates for this test and the significance level at 5%. We use the fanova.tests function from the R package fdANOVA to test for differences in mean functions.

Assessing group differences in correlation functions using the test by Cabassi et al. (2017)

To test for differences in correlation functions between $G$ groups, we proceed in two steps. First, we center the individual trajectories by subtracting them by their group mean trajectories, estimated by taking the sample means of the FPCA-estimated eGFR at each time grid point (years) of the group-specific patients and connecting these sample means to form that group's trajectory. Second, we scale the individual trajectories by dividing them by the square root of the diagonal of the smooth covariance estimates (standard deviations) from our full cohort model. Finally, we apply the multiple-group permutation test developed by Cabassi et al. (2017) to test for differences in covariance functions of our standardized trajectories between $G$ groups. The application of this test is feasible since the covariance of our standardized trajectories is equivalent to the correlation of our un-standardized trajectories.

The global null hypothesis and its alternative are $H_0: \Sigma_1 = \Sigma_2 = \cdots = \Sigma_G$ vs. $H_1: \exists \ u \neq v$ s.t. $\Sigma_u \neq \Sigma_v$, respectively. The permutation test procedure combines the $\frac{G(G-1)}{2}$ partial tests into a single global test by the non-parametric combination algorithm of Pesarin and Salmaso (2010). The partial test statistics are evaluated as the distances between the covariance functions of two



groups, for a pre-defined distance function. Based on the simulation studies and application results of Cabassi et al. (2017), we used the square root distance for $d(\cdot,\ \cdot)$ and the max $T$ combining function for $T_\omega$. Here, the square root distance can be defined as the square root mapping of two covariance operators $\Sigma_1$ and $\Sigma_2$ to the space of Hilbert-Schmidt operators

$$d(\Sigma_1, \Sigma_2) = \left\| \Sigma_1^{1/2} - \Sigma_2^{1/2} \right\|_{HS}.$$

In the case of imbalance between groups, the permutation test procedure would not yield accurate partial p-values when performing permutations for the whole dataset. Therefore, we conducted paired permutations, or permutation tests for each pair of groups independently. Here, the partial tests would be exact, which would allow for two-sample inference of covariance functions. We set 1000 permutation replicates for this test and the significance level at 5%. We use the ksample.perm function from the R package fdcov to test for differences in correlation functions.



Supplementary Table 1: Baseline characteristics of 2641 participants with diabetes in the Chronic Renal Insufficiency Cohort (CRIC) Study.

| | |
|---|---|
| Age (years) | $60.67 \pm 9.48$ |
| Race | |
|    White | 1105 (42) |
|    Black | 1248 (47) |
|    Other | 288 (11) |
| Sex | |
|    Male | 1551 (59) |
|    Female | 1090 (41) |
| Smoked >100 cigarettes | |
|    Yes | 1500 (57) |
|    No | 1141 (43) |
| BMI (kg/m$^2$) | $34.08 \pm 7.77$ |
| HbAlc (%) | $7.62 \pm 1.63$ |
| Diastolic BP (mmHg) | $69.12 \pm 12.36$ |
| Systolic BP (mmHg) | $132.31 \pm 21.76$ |
| Mean Arterial Pressure (mmHg) | $90.19 \pm 13.43$ |
| Serum Creatinine (mg/dL) | $1.74 \pm 0.61$ |
| Urine Creatinine (mg/dL) | $81.1 \pm 51.76$ |
| Continuous Urine PCR (mg/g) | 274.58 [81.92, 1186.44] |
| Predicted Urine ACR (mg/g)* | |
|    < 30 | 965 (37) |
|    30-300 | 768 (29) |
|    > 300 | 908 (34) |
| Baseline eGFR (ml/min/1.73$^2$) | $46.95 \pm 15.23$ |
| Hypertension | |
|    Yes | 2445 (93) |
|    No | 194 (7) |
| ACE Inhibitor or ARB use | |
|    Yes | 2089 (79) |
|    No | 537 (20) |

Values are expressed as mean $\pm$ SD or N (%).
Continuous Urine PCR is expressed as median (IQR, interquartile range) because of its skewed distribution.
*To avoid omitting 842 of the 2641 participants in our sample with missing baseline ACR data, we convert the urine protein-creatinine ratio (PCR) to urine ACR for all participants by employing the crude model by Sumida et al. (2020). These predicted urine ACR values are then used to define our albuminuria groups.
BMI, body mass index; HbA1c, hemoglobin A1c; BP, blood pressure; PCR, protein-creatinine ratio; ACR, albumin-creatinine ratio; eGFR, estimated glomerular filtration rate; ACE, angiotensin-converting enzyme; ARB, angiotensin-receptor blocker.



<u>Supplementary Figure 2:</u> Comparison of estimated individual eGFR trajectories from the full cohort and albuminuria group-specific models for N=5 patients from each of the (a) normal-mild, (b) moderate, and (c) severe albuminuria groups with the group mean trajectories overlaid. The root ACSE (average curve squared error), described in Section 2.3.1, for both the full cohort (FC) and group-specific (gp.) estimated trajectories is displayed for each patient.

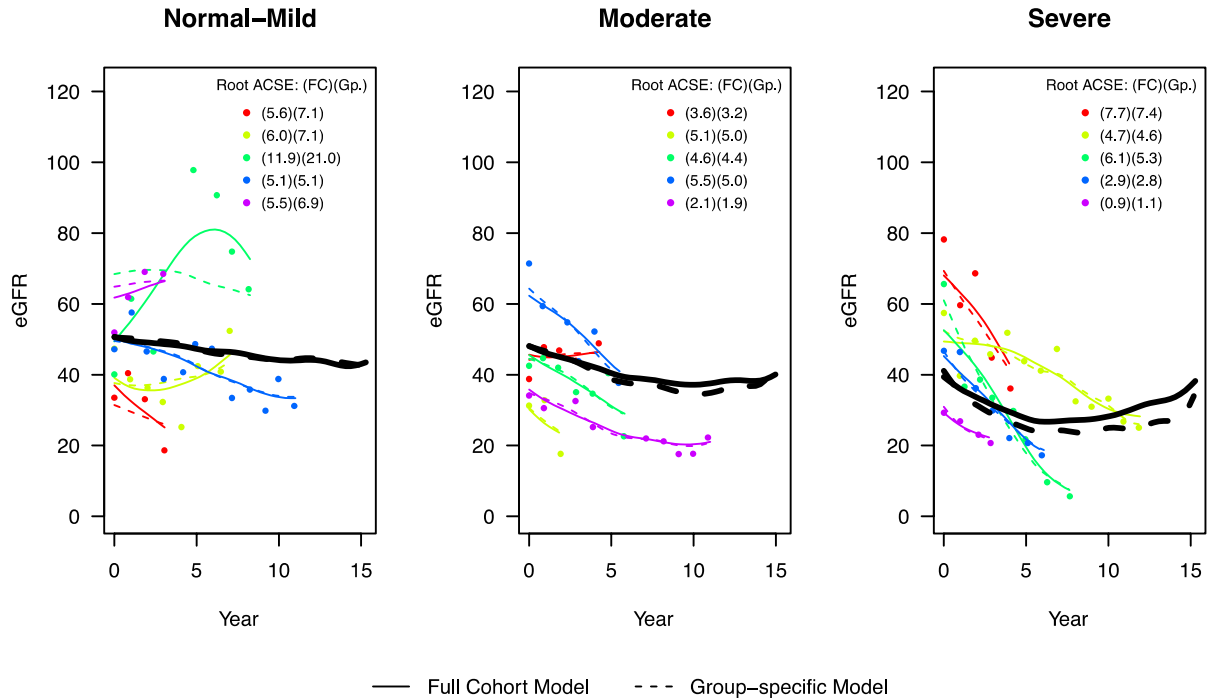